\documentclass[fleqn,twoside]{article}
\usepackage[headings]{espcrc2}
\usepackage{epsfig}
\readRCS
$Id: espcrc2.tex,v 1.2 2004/02/24 11:22:11 spepping Exp $
\newcommand{\AmS}{{\protect\the\textfont2
  A\kern-.1667em\lower.5ex\hbox{M}\kern-.125emS}}

\title{$e^+ e^- \rightarrow 3 ~$jets and event shapes at NNLO} 

\author{G.\ Dissertori\address{Institute for Particle Physics, ETH Z\"urich, CH-8093 Z\"urich, Switzerland},
A.\ Gehrmann-De Ridder\address{Institute for Theoretical Physics, ETH Z\"urich, CH-8093 Z\"urich,
Switzerland},
T.\ Gehrmann\address[UZH]{Institut f\"ur Theoretische Physik,
Universit\"at Z\"urich, CH-8057 Z\"urich, Switzerland},
E.W.N.\ Glover\address[IPPP]{Institute of Particle Physics Phenomenology,
        University of Durham, Durham, DH1 3LE, UK},
G.\ Heinrich\addressmark[IPPP],
H.\ Stenzel\address{II. Physikalisches Institut, Justus-Liebig Universit\"at Giessen,
D-35392 Giessen, Germany}}

\begin{document}

\begin{abstract}
We report on the calculation of NNLO 
corrections to the 3-jet cross section and related 
event shape distributions in electron-positron annihilation. 
The corrections are sizable for all
variables,  however the magnitude of the corrections is substantially 
different for different
observables. We observe that 
inclusion of the  NNLO corrections yields a considerably 
better agreement between theory and experimental data both in shape and 
normalization of the event shape distributions in the region where the perturbative result is expected to hold.
A new extraction of $\alpha_{s}$ using the event shape variables 
up to NNLO yields a considerably better consistency between the observables 
indicating a stabilization of the perturbative corrections at this order.
\end{abstract}

\maketitle

\section{Introduction}
Jet observables in  electron--positron
annihilation play a pivotal role in studying the dynamics of 
the strong interactions, 
described by the theory of quantum chromodynamics (QCD).  
In addition
to measuring multi-jet production rates, more specific information  about the
topology of the events can be extracted  
using variables 
which characterize the hadronic structure of an event. 
With the precision data from LEP and SLC, experimental
distributions for such event shape variables have been extensively  studied 
\cite{Abb05,Heis04} and
have been compared with theoretical calculations based on 
next-to-leading order
(NLO)  parton-level event generator  programs~\cite{Ell81,Kun89}, 
 improved by
resumming kinematically-dominant leading and next-to-leading logarithms
(NLO+NLL)~\cite{Cat93}  and by the inclusion of  
non-perturbative models of power-suppressed hadronisation
effects~\cite{Dok97}. 

Up to now, the precision of the strong coupling constant 
determined from event shape data has been limited  
largely by the scale
uncertainty of the perturbative NLO calculation. 
We report here on the 
first calculation of NNLO corrections to the 3-jet cross section
and related event shape variables. The knowledge of the NNLO corrections 
to the event shape distributions has important phenomenological impact 
on the extraction of $\alpha_{s}$ from LEP data.

\section{The 3-jet cross section at NNLO}

Jets are defined using 
a jet algorithm, which describes how to recombine the momenta 
of all energetic hadrons in an event to form the jets.
These algorithms are used in the experimental analysis 
and in the parton-level event generators to combine particles 
into jets.   
Here we present the first calculation of the NNLO corrections to 
the three-jet production rate at parton-level 
in $e^+e^-$ annihilation using the Durham measure~\cite{Cat91}.

The calculation of the $\alpha_s^3$ corrections for three-jet production
is carried out using a recently developed
parton-level event generator program {\tt EERAD3}
\cite{Gehr07} which contains 
the relevant 
matrix elements with up to five external partons.
Besides explicit infrared divergences from the loop integrals, the 
four-parton and five-parton contributions yield infrared divergent 
contributions if one or two of the final state partons become collinear or 
soft. In order to extract these infrared divergences and combine them with 
the virtual corrections, the antenna subtraction method~\cite{Kos98} 
was extended to NNLO level~\cite{Gehr05} and implemented
for $e^+e^- \to 3\,\mathrm{jets}$~\cite{Gehr08} 
and related event-shape variables~\cite{Gehra07} into {\tt EERAD3}. 
\begin{figure}[t]
\begin{center}
\epsfig{file=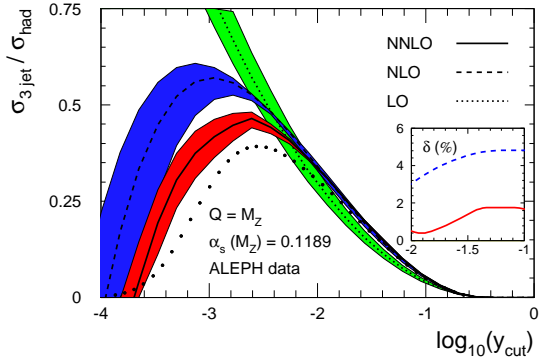,angle=-90,width=7cm}
\end{center}
\caption{Perturbative fixed-order description of the three-jet rate 
in the Durham jet-scheme at $Q=M_{Z}$, compared to data obtained with the ALEPH 
experiment \protect\cite{Heis04}.}
\label{fig:3j}
\end{figure} 

Figure~\ref{fig:3j} displays the three-jet rate 
at LEP1 energy $Q=M_Z$ as function of the jet 
resolution $y_{cut}$ at LO, NLO, NNLO. 
The theoretical uncertainty band is
defined by varying the renormalization scale $\mu$
in the coupling constant in the interval $M_Z/2 < \mu < 2\,M_Z$, 
and the average value~\cite{Beth07} $\alpha_s(M_Z) = 0.1189$
is  used, consistently evolved to other scales at each order.

Since the error band in the region $10^{-1}> y_{cut} > 10^{-2}$ 
is barely visible in the plot, we 
display the relative theoretical uncertainty 
\begin{displaymath}
\delta = \frac{\max_{\mu} (\sigma(\mu)) - \min_\mu (\sigma(\mu)) }
{2 \sigma (\mu = M_Z)}
\end{displaymath}
at NLO and NNLO 
as an inset. The uncertainty on the LO calculation is constant at 10.2\%. 

As can be seen from the plot,
the theoretical uncertainty is lowered considerably compared to NLO. 
Especially in the region $10^{-1}> y_{cut} > 10^{-2}$, which is relevant 
for precision phenomenology, one observes a reduction by almost a factor 
three, down to below two per cent relative uncertainty. 

For large values of $y_{cut}$, $y_{cut} > 10^{-2}$, the NNLO corrections 
turn out to be very small, while they become substantial for medium and 
low values of $y_{cut}$. The maximum of the jet rate is shifted towards 
higher values of $y_{cut}$ compared to NLO, and is in better  
agreement with the experimental observation.

The fixed-order theoretical predictions 
for the three-jet rate become negative for small values of 
$y_{cut}$, where fixed order perturbation theory is not applicable due to the 
emergence of large logarithmic corrections at all orders which 
require resummation~\cite{Cat91,Ban02}. 
We therefore restrict our comparison to $y_{cut} > 10^{-4}$.
Even with this restriction, at low jet resolution, 
the fixed-order NNLO description lies above the data.
The theoretical parton-level prediction 
is compared however to hadron-level data, thereby neglecting hadronisation 
corrections, which may account for part of the discrepancy.

The total hadronic cross section consists of the sum over
all jet multiplicities. At ${\cal O}(\alpha_s^3)$, this sum 
runs from two-jet through to five-jet final states, 
such that the corresponding fractional jet rates must add to unity. Consequently, our 
calculation yields the N$^3$LO 
expression for $e^+e^- \to 2$~jets as a by-product.

\begin{figure*}[t]
\begin{center}
\parbox{15.8cm}{\epsfig{file=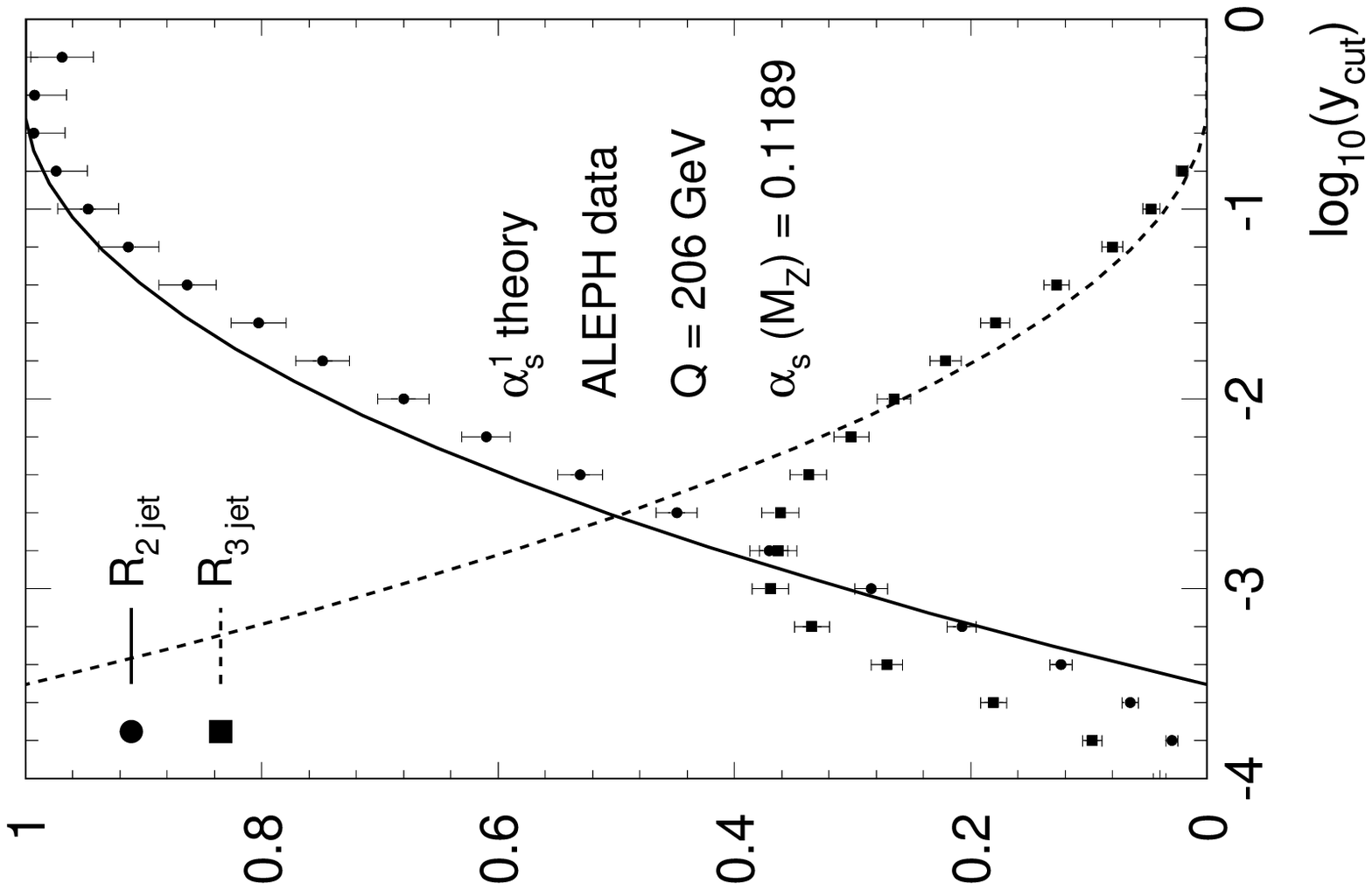,angle=-90,width=5.0cm}
\epsfig{file=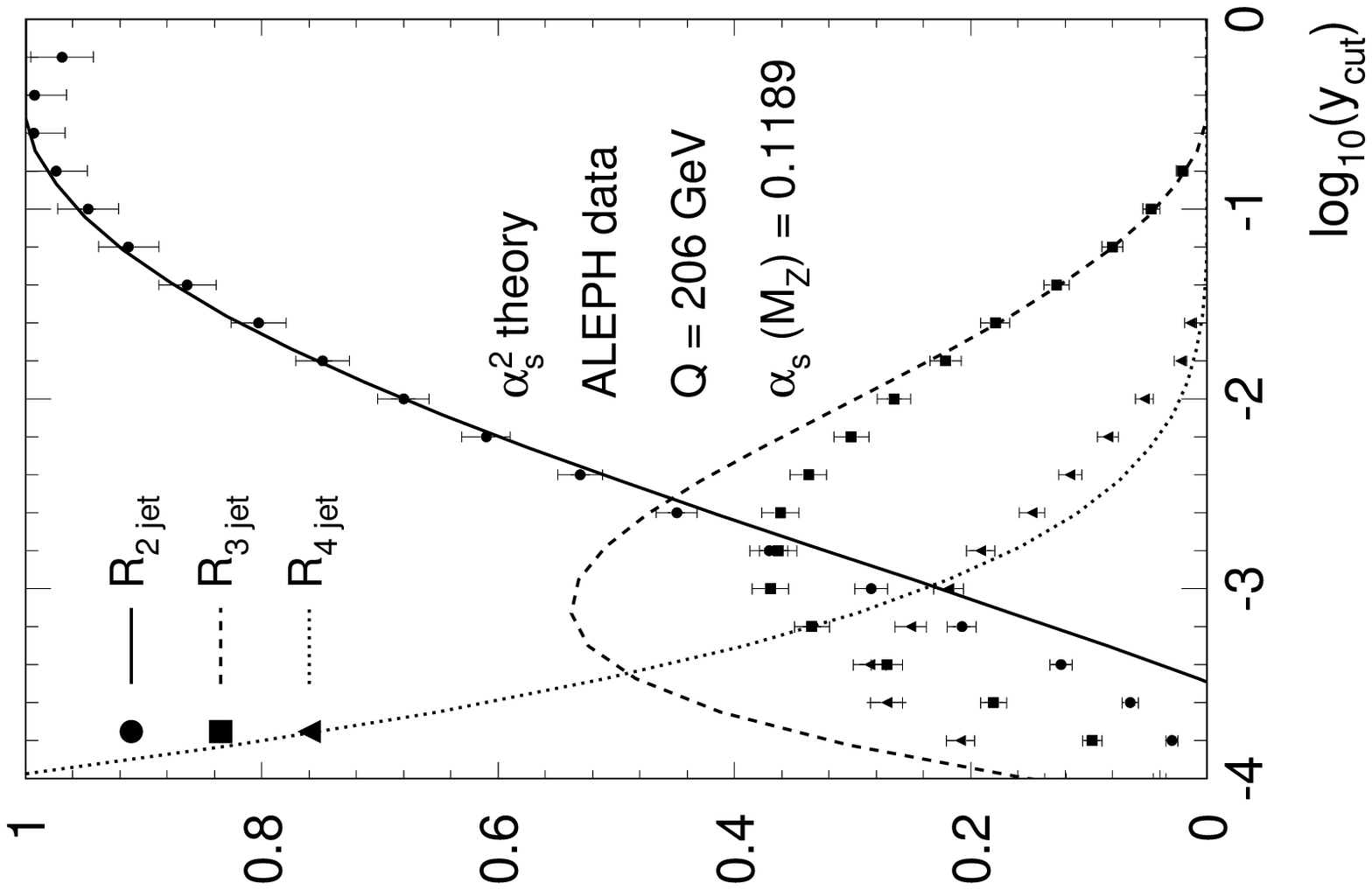,angle=-90,width=5.0cm}
\epsfig{file=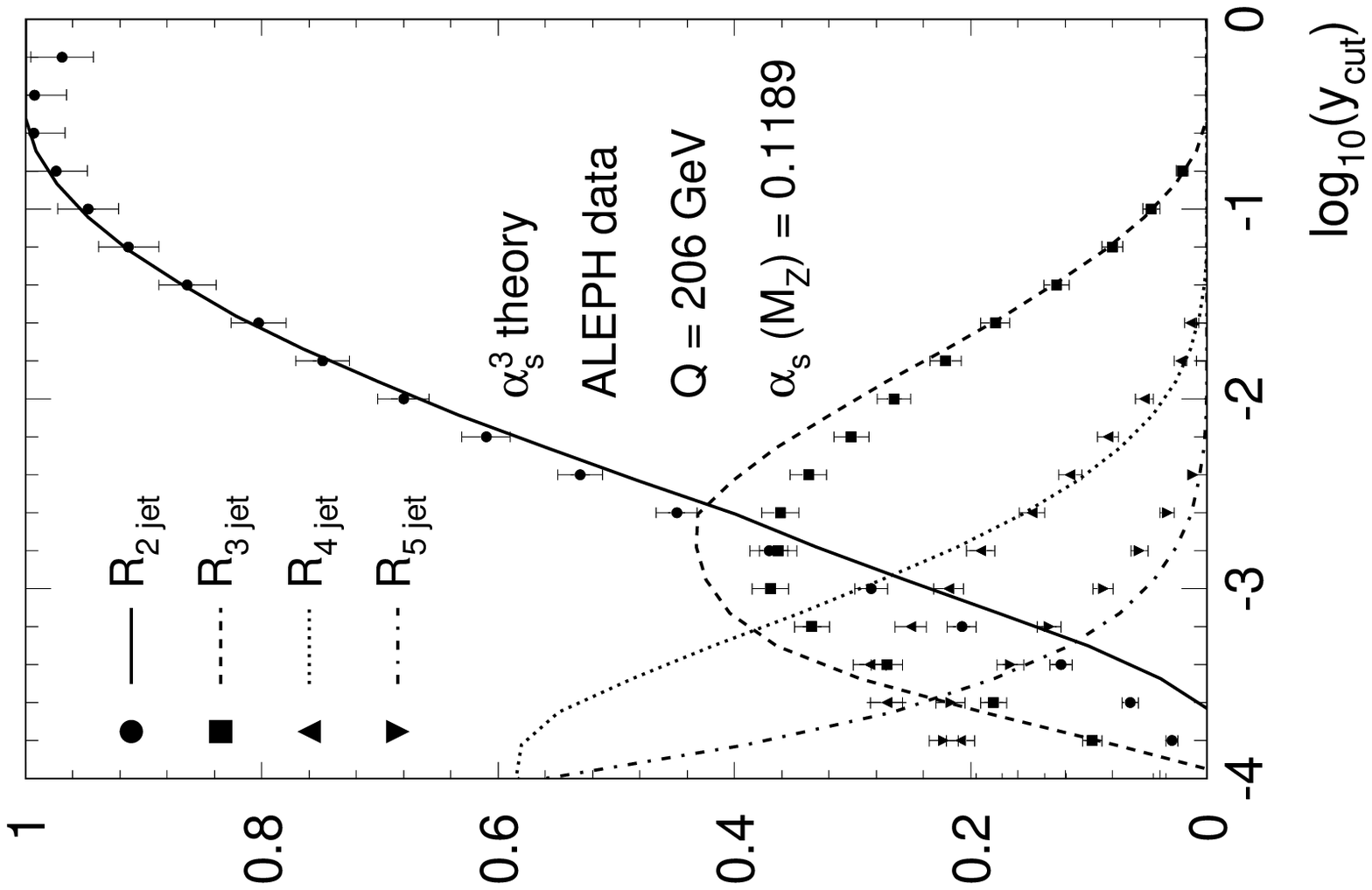,angle=-90,width=5.0cm}}
\end{center}
\caption{Jet rates at first, second and third order in the 
strong coupling constant, compared to data 
from ALEPH~\protect{\cite{Heis04}}.
The rates are normalized to the total
hadronic cross section at that order.
\label{fig:rates}}
\end{figure*}
Figure~\ref{fig:rates} shows the parton-level theoretical predictions for the 
jet fractions at first, second and third
order
in the  strong coupling constant, compared to experimental hadron-level 
data from ALEPH~\cite{Heis04}. 

By comparing the three plots, we observe that the 
agreement for each of the jet rates becomes systematically 
better as the order of perturbation theory
increases.  At each order a new multi-jet channel opens up, e.g. the five-jet
rate at ${\cal O}(\alpha_s^3)$, which is positive definite and essentially
monotonically increasing as $y_{cut}$ decreases.  Since all jet rates are normalized
to unity, the new five-jet channel has the effect of reducing the contribution
to the two-jet, three-jet and four-jet rates, in the region of $\log_{10}(y_{cut})$
where the five-jet rate contributes.  
One very clear effect is to cause the turnover in the four-jet rate (which is not present at ${\cal O}(\alpha_s^2)$). 
A second effect is to add more structure to the shape of the two- and three-jet
rates, which lie much closer to the data for $\log_{10}(y_{cut}) < -2.5$.
Of course, the effect of the higher order corrections also extends to larger values of $y_{cut}$ 
and, by adding more structure to the theoretical prediction, one obtains a better description of the data.

\section{Event shape variables}

In order to characterize hadronic final states in electron-positron
annihilation, a variety of event shape variables have been proposed in 
the literature. For a review see e.g.~\cite{Ell03,Jon03}. 
In our study we considered only variables for 
three-particle final states which are thus closely related to three-jet 
final states.
Among these event  shapes,
six variables
were studied in great detail \cite{Abb05}: the thrust $T$, the
normalized heavy jet mass $\rho$, 
the wide and total jet
broadenings $B_W$ and $B_T$,  
the $C$-parameter and the transition from three-jet to 
two-jet final states in the Durham jet algorithm called $Y_3$.

The perturbative expansion for the distribution of a 
generic observable $y$ up to NNLO at $e^+e^-$ centre-of-mass energy $\sqrt{s}$, 
for a renormalization scale $\mu^2$ involves perturbative coefficients 
\cite{Gehra07}
which only depend on the event shape variable $y$ itself and the strong
coupling constant $\alpha_s$. 
Those coefficients are computed by a fixed-order 
parton-level calculation, which includes final states with three partons 
at LO, up to four partons at NLO and up to five partons at NNLO. 

The precise size and shape of the NNLO corrections depend on the observable 
in question. Common to all observables is the divergent behaviour of 
the fixed-order prediction in the two-jet limit (corresponding to
small values of the event shape variable $y$), where soft-gluon effects 
lead to an enhancement of 
the fixed-order coefficients by powers of $\ln(1/y)$.
In order to obtain reliable predictions
in the region of $y \ll 1$ it is necessary to resum entire sets 
of logarithmic terms at all orders in $\alpha_s$. 
A detailed description of the predictions at next-to-leading-logarithmic 
approximation (NLLA) can be found in Ref.~\cite{Jon03}. 

\begin{figure}[t]
\begin{center}
\epsfig{file=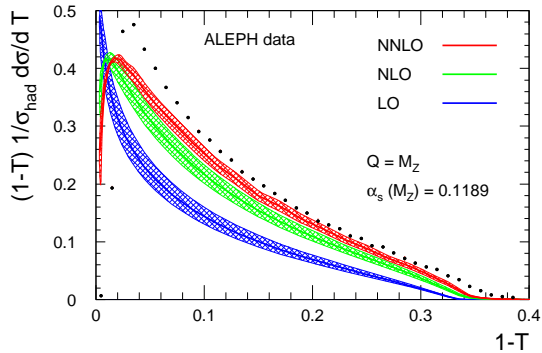,angle=-90,width=7cm}
\end{center}
\caption{Thrust distribution at $Q= M_Z$ at LO (blue), NLO (green) and NNLO
(red). The solid lines represent the prediction for renormalisation scale 
$\mu=Q$ and $\alpha_s(M_Z) = 0.1189$, while the
shaded region shows the variation due to varying
the renormalisation scale between $\mu=Q/2$ and $\mu = 2 Q$.
The data is taken from \protect{\cite{Heis04}}.\label{fig:thrust}}
\end{figure}

For several 
event shape variables 
(especially $T$ and $C$) the full kinematical range is not yet realised 
for three partons, but attained only in the multi-jet limit. 
For the thrust distribution $1-T$, as seen in  Fig.~\ref{fig:thrust}, the
multi-jet limit corresponds to $1-T>0.5$. 
Consequently, the 
fixed-order description is expected to be reliable in a restricted 
interval bounded by the two-jet limit on one side and the multi-jet 
limit on the other side. 

In the intermediate region, we observe that 
inclusion of  NNLO corrections (evaluated at the $Z$-boson mass, and 
for fixed value of the strong coupling constant) typically increase 
the previously available NLO prediction. 
The magnitude of this increase differs considerably between 
different observables\cite{Gehra07}, 
it is substantial for $T$ (18\%), $B_T$ (17\%)  and 
$C$ (15\%), moderate for $\rho$ and $B_W$ (both 10\%) and small for 
$Y_3$ (6\%). For all shape variables, we observe that the renormalization
scale uncertainty of the NNLO prediction is reduced by a factor 2 or more
compared to the NLO prediction. 
We observe that 
the NNLO prediction describes the shape of the measured event shape 
distributions over a wider kinematical range than the NLO prediction, both 
towards the two-jet and the multi-jet limit. 

\section{Determination of the strong coupling constant}

Using the newly computed NNLO corrections to event shape variables, we
performed \cite{Dis08} 
a new extraction of $\alpha_s$ from data on the standard set of 
six event shape variables measured 
by the ALEPH\ collaboration \cite{Heis04}
at centre-of-mass energies of 91.2, 133, 161, 172, 183, 189, 200 and 206 GeV.
The combination of 
all NNLO determinations from all shape variables  yields 
\begin{eqnarray*}
 \lefteqn{   \alpha_s(M_Z) = 0.1240 \;\pm\; 0.0008\,\mathrm{(stat)}} \\
     			&&		 \;\pm\; 0.0010\,\mathrm{(exp)}
                                   \;\pm\; 0.0011\,\mathrm{(had)} \\
  &&                                   \;\pm\; 0.0029\,\mathrm{(theo)} .
 \end{eqnarray*}

We observe a clear improvement in the fit quality when going to
NNLO accuracy. Compared to NLO, the value of $\alpha_s$ is lowered 
by about 10\%, but still higher than for NLO+NLLA~\cite{Heis04},
which shows the obvious need for a matching of NNLO+NLLA for an even 
more precise 
result. Work is in progress in this direction \cite{Gehrt08}. 
 
\begin{figure}[t]
\begin{center}
\epsfig{file=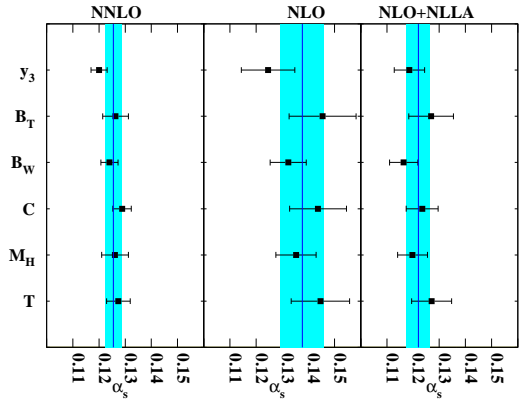,width=7cm}
\end{center}
\caption{\small The measurements of the strong coupling constant $\alpha_s$
for the six event shapes, at $\sqrt{s}=M_{z}$, when using QCD predictions 
at different approximations in perturbation theory. The blue band indicates
the uncertainty due to renormalisation scale variation in each 
theoretical description. }
\label{fig:scatter}
\end{figure}

As can be seen in Figure~\ref{fig:scatter}, the scatter among the
$\alpha_s$-values extracted from different shape variables is 
lowered considerably when going to NNLO accuracy
 and the theoretical uncertainty is decreased by 
a factor 2 (1.3) compared to NLO (NLO+NNLA). 
The different sizes of the NNLO corrections  
for different observables is responsible for these large improvements.  
One infers that the scatter present at next-to-leading order was largely due 
to missing higher order perturbative corrections.

\section{Outlook}

Our results for the NNLO corrections to the 3-jet cross section and related 
event shape distributions in electron-positron annihilation  open up a 
whole new
range of possible comparisons with the LEP data.  The potential of these studies
is illustrated by comparisons of the NNLO fixed order results with jet and
event-shape data from ALEPH. The corrections are sizable for all variables, but
yield a considerably  better consistency between the observables  indicating a
stabilization of the perturbative corrections at this order. A fit to event
shape data yielded a new determination of $\alpha_s$.    We anticipate a further
improvement by matching of the fixed order NNLO calculation with NLLA
resummations~\cite{Gehrt08}. 

\section*{Acknowledgments}

This work was supported by the Swiss National Science Foundation  (SNF) under
contracts PP002-118864 and 200020-117602,  by the UK Science and Technology
Facilities Council and  by the European Commission's Marie-Curie Research
Training Network under contract MRTN-CT-2006-035505 ``Tools and Precision
Calculations for Physics Discoveries at Colliders''.

Furthermore this work has made use of the resources provided by the
Edinburgh Computer and Data Facility (ECDF).

\end{document}